\documentclass[a4paper]{article}
\makeatletter
\let\@fnsymbol\@arabic
\makeatother
\usepackage[pages=all, color=black, position={current page.south}, placement=bottom, scale=1, opacity=1, vshift=5mm]{background}
\SetBgContents{
}      

\usepackage[margin=1in]{geometry} 

\usepackage{amsmath}
\usepackage{amsthm}
\usepackage{amssymb}

\usepackage{caption}
\usepackage{subcaption}

\usepackage{appendix}
\usepackage{etoolbox}
\usepackage{authblk}

\BeforeBeginEnvironment{appendices}{\clearpage}

\usepackage[utf8]{inputenc}
\usepackage{hyperref}
\hypersetup{
	unicode,
	pdfauthor={Author One, Author Two, Author Three},
	pdftitle={A simple article template},
	pdfsubject={A simple article template},
	pdfkeywords={article, template, simple},
	pdfproducer={LaTeX},
	pdfcreator={pdflatex}
}

\usepackage[sort&compress,numbers,square]{natbib}
\usepackage{graphicx, color}
\graphicspath{{fig/}}

\usepackage{algorithm, algpseudocode} 
\usepackage{mathrsfs} 

\theoremstyle{plain}

\theoremstyle{definition}

\title{CompanyName2Vec: \linebreak Company Entity Matching Based on Job Ads}
\author[*]{Ran Ziv\thanks{ran.ziv@post.idc.ac.il}}
\author[*]{Ilan Gronau\thanks{ilan.gronau@post.idc.ac.il}}
\author[**]{Michael Fire\thanks{mickyfi@bgu.ac.il}}
\affil[*]{Efi Arazi School of Computer Science, Reichman University}
\affil[**]{Department of Software and Information Systems Engineering, Ben-Gurion University}

\begin{document}
	\maketitle
	\begin{abstract}
		
        Entity Matching is an essential part of all real-world systems that take in structured and unstructured data coming from different sources. Typically no common key is available for connecting records. Massive data cleaning and integration processes require completion before any data analytics, or further processing can be performed. Although record linkage is frequently regarded as a somewhat tedious but necessary step, it reveals valuable insights, supports data visualization, and guides further analytic approaches to the data. Here, we focus on organization entity matching. We introduce CompanyName2Vec, a novel algorithm to solve company entity matching (CEM) using a neural network model to learn company name semantics from a job ad corpus, without relying on any information on the matched company besides its name. Based on a real-world data, we show that CompanyName2Vec outperforms other evaluated methods and solves the CEM challenge with an average success rate of 89.3\%.
        \\
        
		\noindent\textbf{Keywords:} Entity Matching, Organization Name Matching, LSTM, CompanyName2Vec
	\end{abstract}

	\section{Introduction}
        
        Enterprise Business intelligence systems have emerged as a disruptive technology and innovative solution to the global economy \cite{scholes2011lessons}. Business intelligence became an emerging and fast-growing field in the past years \cite{liebowitz2014business}. Big data and its emerging technologies, including business intelligence and big data analytics systems, have become a mainstream market adopted broadly across industries, organizations, and geographic regions to facilitate data-driven decision making and significantly affect the way that decision-makers, such as CEOs, operate and run their business \cite{liebowitz2014business, sun2018big}. One of the fundamental functionalities a business intelligence system must have is the ability to integrate many data sources \cite{liebowitz2014business, gschwind2019fast}. The integration of multiple sources usually requires linking between the significant entities that exist across data sources and systems. Usually, those entities function as a ``primary key'' in each system \cite{gschwind2019fast}. The technique used for performing such linkage is commonly referred to as ``Record Linkage,'' ``Data Deduplication,'' ``Object Matching,'' or ``Entity Matching'' \cite{gschwind2019fast, elmagarmid2006duplicate}. 
        
        Entity Matching (EM) is a fundamental task in data integration scenarios. EM is the task to identify semantically equivalent entities referring to the same real-world object (e.g., persons, products, companies) within one data source or between different sources. EM is also a core technique for data cleaning \cite{elmagarmid2006duplicate, rahm2000data} and data integration \cite{vesset2013worldwide, paganelli2019tuner}. Accurate and fast entity matching has huge practical implications in a wide variety of commercial, scientific and security domains \cite{getoor2012entity}. EM solutions can be divided into three separate groups: supervised, unsupervised, and semi-supervised approaches \cite{veldman2009matching, chaudhuri2005robust}. Supervised approaches require labeled training sets or predefined thresholds on which to base their decisions. Unfortunately, in most real-world cases, the variety of patterns that can be observed are not feasible to be captured in a training set. Therefore, these solutions are quite limited. Unsupervised approaches help find previously unknown patterns in a dataset without pre-existing labels and are based on clustering algorithms that group together items with high similarity \cite{bousquet2011advanced}. Semi-supervised approaches fall between the unsupervised approach (with no labeled training data) and a supervised approach (with only labeled training data). Semi-supervised approaches are based on a small set of labeled data and a large set of unlabeled data. 
        
        Many enterprise business intelligence applications require the integration of multiple data sources. In such applications, a company name is one of the most crucial entity attributes to be linked \cite{gschwind2019fast}. A company usually exists in many of the enterprise systems. For example, a company can represent a potential customer, a sales lead in the sales system, an existing customer in the customer relationship management system, a supplier in the logistics system, etc. Although a company name can be noisy, in most cases, this is one of the most potent properties for company entity matching (CEM) due to the lack of unique shared identifiers across datasets \cite{gschwind2019fast}. 

        Determine whether a new company name is in a database, and if so, which existing record it refers to is a typical problem business intelligence applications need to solve given records indexed by company names and a new company name.
        This problem is an instance of entity matching problem. It is a challenging problem because people do not consistently use the official name, but use abbreviations, synonyms, different order of terms, different spelling of terms, short form of terms, and the name can contain typos or spacing issues.

        Many of the existing methods use more company properties other than the company name, such as location, primary phone number, industry, website URL, and more to improve CEM results. Such company information is usually costly to acquire on a large scale. The largest public source for such company information is DBpedia \cite{dbpedia}.  DBpedia contains about 65,000 company entities worldwide, derived from the English version of Wikipedia \cite{gschwind2019fast}. There are several companies worldwide that hold a much larger dataset of company information, such as Dun and Bradstreet and Infogroup \cite{jones2017step}. These companies do not provide the complete dataset but provide a limited paid service for company information enrichment, mainly for sales and marketing purposes. 
        
        In this study, we developed CompanyName2Vec (see Section~\ref{method}), a novel algorithm to solve CEM using a neural network model to learn company name semantics from a job ad corpus. Once trained, such a model can suggest synonymous company names. As the name implies, CompanyName2Vec represents each different company name as a vector. The vectors are chosen carefully such that a simple mathematical function, like cosine similarity between vectors, indicates the level of semantic similarity between the company names represented by those vectors.
        We used CompanyName2Vec for CEM. For this purpose, we created a real-world dataset that consists of the largest companies in the US and their synonyms. We used this dataset to evaluate the CompanyName2Vec algorithm and compared it with other known entity matching methods.
        
        To develop the CompanyName2Vec algorithm, we needed a relatively large corpus of company names. Job ads consisting of company names that are available publicly on a large scale but are very noisy (see Section~\ref{method-hiring company names fingerprinting}) due to how job ad distribution works. Companies distribute their jobs to multiple job boards and sometimes to recruitment and staffing agencies to reach as many job seekers as possible. In many cases, the process for posting a job ad on job boards requires the employer to manually type the job information, including company name, location, title, description, and more, which causes some inconsistency and duplication of job ads. For example, Tesla’s job ad can be posted with several employer names like tesla; Tesla, Inc.; Tesla Motors; Tesla Motors, Inc.; and Ursus, which is a staffing and recruitment agency. 
        CompanyName2Vec leverages the availability of job ads publicly (see Section~\ref{Experimental Study - dataset}) and their inconsistency to learn company name semantics (see Section~\ref{method-company name embedding}). 
        
\begin{figure*}[ht]
            \centering
            \includegraphics[scale=0.6]{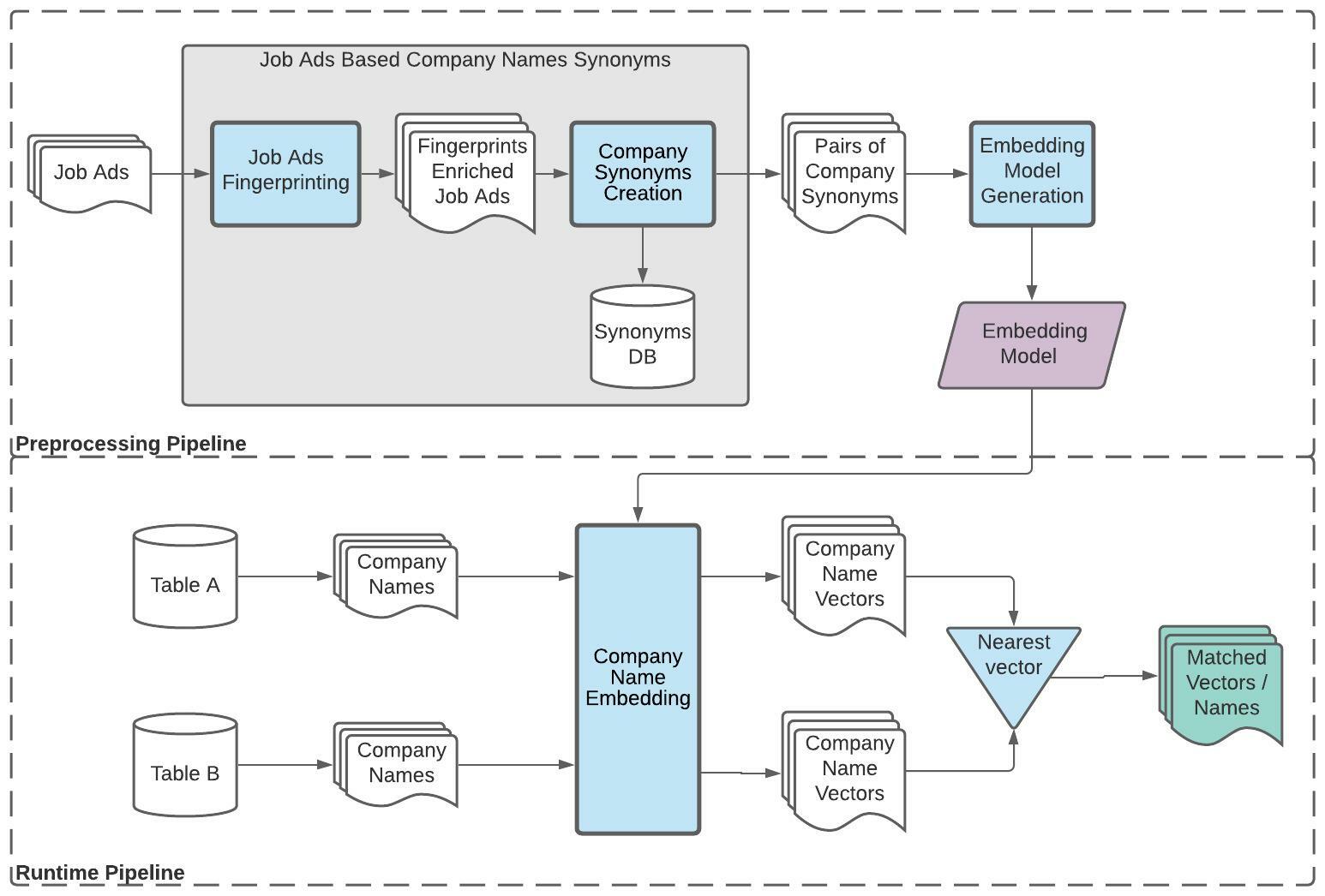}
            \caption{Flowchart of the implemented methodology.}
            \label{fig:methodworkflow}
        \end{figure*}

        The method developed consists of the following steps (see Figure \ref{fig:methodworkflow}): First, we collected a large dataset of job ads (see Section~\ref{Experimental Study - dataset}). Second, we used a fingerprinting technique and created a list of company names that posted the same job (see Section~\ref{method-hiring company names fingerprinting}). Additionally, we cleaned the data to filter out non-valid job ads and company names to create a ground truth dataset.
        Next, we pre-processed this cleaned dataset and created train and test sets (see Section~\ref{method-company name embedding}). 
        Afterward, we trained a neural network model to learn company names semantics and transform company names into a vector representation and used cosine distance for measuring the similarity between different vectors, where two company names that represent the same company will have similar vectors (see Section~\ref{method-company name embedding}). For example, employers write their company name in many ways, sometimes ignoring common suffixes, like Inc., and LLC, which are used to define the company legal entity type but are less important when posting a job ad or tagging the company name in a social media post. 
        Lastly, we created a real-world dataset of the largest 1,000 companies in the US and their synonyms (see Section \ref{Experimental Study - dataset}). We used this dataset to evaluate our method's performance by measuring the success rate at $k$ of matching company synonyms with their mapped canonical company name (see Section \ref{Evaluation Process and Performance Metrics}). 
        
        This study's main contributions are twofold:
        
        \begin{itemize}
            \item  \textit{CompanyName2Vec,} a novel algorithm to create a mathematical representation for a company name that holds the naming semantics, which enables it to measure similarity between any given company names without relying on any information on the matched company besides its name  (see Section~\ref{method}).
            \item \textit{An end-to-end, highly scalable, enterprise-grade system} that uses CompanyName2Vec algorithm to suggest company name synonyms to solve company entity matching (see Section~\ref{Experimental Study}).
        \end{itemize}
        
        The remainder of the paper is organized as follows: In Section~\ref{related work}, we provide
        an overview of related studies. Next, in Section~\ref{method}, we present the methods used to create the CompanyName2Vec embedding model. Afterwards, in Section~\ref{Experimental Study}, we describe the experimental study we conducted to construct the model and evaluate it.
        Subsequently, in Section~\ref{experimental-study-results}, we present the obtained results. Then, in Section~\ref{sec:discuss}, we discuss the obtained results. Lastly, in  Section~\ref{sec:conclusions}, we present our conclusions from this study and offer future research directions.

	\section{Related Work} \label{related work}
        
        Entity Matching (EM) is an important task, which many researchers in the past have addressed \cite{elmagarmid2006duplicate, getoor2005link, benjelloun2009swoosh, brizan2006survey}. 
        In this study, we addressed a particular case of EM, namely, the matching of company names across multiple datasets and systems. 
        
        In Section~\ref{related work - entity matching}, we present the current approaches concerning EM. In section~\ref{related work - Distance Metrics} we describe several distance metrics methods and techniques. In section~\ref{related work - fingerprinting}, we describe methods for document fingerprinting. Lastly, in section~\ref{related work - embedding}, we explain embedding with an emphasis on text embedding. 
        
        \subsection{Entity Matching} \label{related work - entity matching}
        EM addresses matching entities between different data sources or deduplication of entities within a single source. EM uses algorithms to both detect duplicates in data and resolve them. As described above, EM solutions can be divided into three separate groups: supervised, unsupervised, and semi-supervised approaches. In the following sections, we describe the different methods for EM.
        
        \subsubsection{Supervised Approaches}
        The supervised approaches are based on training data in the form of record pairs, pre-labeled as match or not match. In 1969, Ivan Fellegi and Alan Sunter \cite{fellegi1969theory} denote two classes: A class \textit{M}, which represents matches, and a class \textit{U}, which represents non-matches. There are three main approaches: Rule-based, learning-based, and distance-based.
        
        \textit{Rule-Based.} One way of performing entity matching is by setting a set of rules that identify the conditions that would make two records be considered as matched. For instance, matching a customer entity which includes its name and full address, it could be of the form: 
        $ (name, edit, =, 1) \land (address, edit, >, 0.7)$, which indicates that two customer records would match if their names are fully matched and their address and city attributes are more than 70\% similar. 
        
        \textit{Learning-Based.} Learning-based approaches \cite{elmagarmid2006duplicate} use training sets that consist of pairs labeled as Match or Non-match. Each pair includes a comparison vector that represents the comparable attributes of the two items in the pair. Assuming the density function for the classes M and U are different, the EM problem can be treated as a Bayesian inference problem.
        
        \textit{Distance-Based.} Distance-based approaches \cite{elmagarmid2006duplicate} theoretically do not need labeled data. In these approaches, a distance metric is defined between data items. A decision is made based on whether or not this pair is a match or not based on the distance between two items and a threshold. This threshold can be set by making a reliable estimation. A good threshold can improve the results. Therefore using a training set for setting this threshold might be desirable. 
        
        \subsubsection{Unsupervised Approaches}
        The idea behind unsupervised approaches is that similar comparison vectors correspond to the same class \cite{elmagarmid2006duplicate}. Unsupervised learning for EM has its roots in the probabilistic model proposed in 1969 by Ivan Fellegi, and Alan Sunter \cite{fellegi1969theory}. When there is no training data to compute the probability estimates, it is possible to use variations of the Expectation-Maximization algorithm \cite{dempster1977maximum} to better identify appropriate clusters in the data.
        
        \subsection{Distance Metrics} \label{related work - Distance Metrics}
        A common source for mismatches in database entries is the typographical variation of string data. One of the methods to deal with typographical variations is to measure the similarity between two different strings. In this section, we describe string matching techniques that have been applied in EM.
        
        \subsubsection{Character-Based Similarity Metrics} \label{Character-Based Similarity Metrics}
        The character-based similarity metrics method is designed to  manage typographical errors efficiently. A popular character-based method is the Levenshtein distance (also referred to as Edit Distance) \cite{navarro2001guided}. This method measures similarity between two strings by counting the minimum number of operations required to transform one string into the other, including insertion, deletion or substitutions. Damerau-Levenshtein distance \cite{navarro2001guided, damerau1964technique} is a variation of the Levenshtein distance where a transposition of two characters is also considered to be an elementary edit operation in addition to insertion, deletion, and substitution. Jaro distance \cite{yancey2005evaluating} is based on the number of common characters and the number of transpositions in two strings. Winkler distance \cite{yancey2005evaluating} is a variation of the Jaro distance and gives higher scores to strings that share the same prefix. 
        
        In terms of company names, Levenshtein Distance seems to work well for a single word or a much longer text, but not for just for a few words \cite{zhang2017research}, as in our use case. For example, ``New York Yankees'' and ``Yankees'' are clearly referring to the same company name, but ``New York Mets'' and ``New York Yankees'' are clearly referring to different ones. Yet, the score of the ``wrong'' match is higher than the ``right'' one. 
        For such cases, there is a heuristic called ``best partial'' \cite{fuzzywuzzypython}, which uses partial matching logic. This logic is implemented in the Fuzzy-Wuzzy package \cite{fuzzywuzzypython}, which is based on Levenshtein Distance \cite{navarro2001guided} to calculate the differences between two different strings. Given two strings $X$ and $Y$, let the shorter string $X$ be of length $m$. It finds the ratio similarity measure between the shorter string $X$ and every substring of length $m$ in the longer string $Y$, and returns the maximum of those similarity measures. 
        So in the case of ``Yankees'' and ``New York Yankees'' the score will be higher than the score of ``New York Mets'' and ``New York Yankees'' (see Table~\ref{table:CompanyNamesSynonymsRatio}), since the substring ``Yankees'' is wholly contained in the string ``New York Yankees.''

        \begin{table}[htb]
        
        \centering
        \caption{Company Names Synonyms Ratio Similarity vs. Partial Ratio Similarity}
        \label{table:CompanyNamesSynonymsRatio}
        \begin{tabular}{||c c c c||} 
         \hline
         \textbf{String X} & \textbf{String Y} & \textbf{Ratio} &   \textbf{Partial Ratio}\\ 
                  &          & \textbf{Similarity} &  \textbf{Similarity}  \\
         \hline\hline
         YANKEES & NEW YORK & $61\%$ & $100\%$ \\
                 & YANKEES &  &  \\ 
         \hline
         NEW YORK  & NEW YORK  & $76\%$ & $69\%$ \\
         METS     & YANKEES &  &  \\
         \hline
         NEW YORK  & NEW YORK  & $96\%$ & $92\%$ \\
         METS & MEATS & &  \\ 
         \hline
        \end{tabular}

        \end{table}

        \subsubsection{Token-Based Similarity Metrics}
        The method of Character-based similarity metrics measures similarity depending only on the appearance and sequence of characters, while token-based similarity metrics first tokenize two strings into sets of tokens (words) and only then compute the similarity between the two sets. Token-based similarity metrics are robust in measuring the similarity of full names, for example, where the order of the first name and last name may change from one string to another. 
        A standard method for Token-based similarity metrics is \textit{TF-IDF} (Term Frequency / Inverted Document Frequency) \cite{cohen1998integration}, a numerical statistic method that intends to reflect how important a word is to a document in a collection of documents. 
        \textit{Cosine similarity} \cite{cohen1998integration} measures the similarity of strings by transforming words into vectors, where the frequency of a word is a dimension in the vector. Cosine similarity measures the similarity between two strings by measuring the angle between the vectors. Token-based similarity metrics methods do not take misspellings into account. Therefore those methods are usually used together with character-based methods to determine whether two tokens are similar enough \cite{cohen1998integration}. 
        
        \subsubsection{Phonetic Similarity Metrics}
        Strings may be phonetically similar even if they are not similar at the character or token level. Different from character-based and token-based similarity metrics, phonetic similarity metrics are limited to a string-based representation. An example of this is the name Claire. It has two alternatives, Clare and Clair, which are both pronounced the same.  Soundex \cite{holmes2002improving, lait1996assessment}, for example, is one of the best known phonetic encoding algorithms for indexing names by sound as pronounced in English. Soundex keeps the first letter and converts the rest of the string into numbers according to a phonetic encoding table. Additional string comparison methods can be found in a thorough survey written by Peter Christen \cite{christen2006comparison}.

        \subsection{Document Fingerprinting} \label{related work - fingerprinting}
        Among digital data, documents are the easiest to copy and remove any signatures or fingerprints embedded, which make the pirating the hardest to detect \cite{elbegbayan2005winnowing}. Anyone can retype a document or copy a part of it. Document fingerprinting is concerned with accurately identifying and copying, including small partial copies, within large sets of documents \cite{elbegbayan2005winnowing}.
        Our study uses fingerprinting to find hiring company synonyms in the job ads dataset by calculating a job ad fingerprint and looking for hiring company names with the same fingerprint. A Checksum, for example, is a small digest of the entire document. This method is simple and sufficient for detecting exact copies. Still in some cases, as well as for our study, there is a requirement in a method that is more local and detects partial copies like the Winnowing algorithm \cite{schleimer2003winnowing}, which selects the q-gram whose hash value is the minimum within a sliding window of q-grams. Winnowing is used for text clustering \cite{parapar2008winnowing}, and detecting plagiarism, like the comparison Agung Toto Wibowo published in 2013 \cite{wibowo2013comparison} which detects plagiarism fraud on Bahasa Indonesia documents, or Moss by Alex Aiken \cite{mossatstanford} which is a system for measuring software similarity, used by Stanford university for detecting plagiarism in programming classes.

        \subsection{Text Embedding} \label{related work - embedding}
        In this study, we used embedding to create a mathematical representation for a company name that holds the naming semantics. 
        When some object $x$ is said to be embedded in another object $y$, the embedding is given by some injective and structure-preserving map $f : X \to Y$. 
        We used text embedding, similarly to what Yoshua Bengio proposed in 2001~\cite{bengio2001neural}, in the form of a feed-forward neural network language model. Modern methods use a simpler and more efficient neural architecture to learn word vectors, like word2vec \cite{mikolov2013efficient, mikolov2013distributed}, and GloVe \cite{pennington2014glove}, based on objective functions that are explicitly designed to produce high-quality vectors.
        Neural embedding learned by these methods have been applied in a myriad of NLP applications, including initializing neural network models for objective visual recognition \cite{frome2013devise}, or machine translation \cite{li2014neural, zhang2014bilingually}, as well as directly modeling word-to-word relationships \cite{mikolov2013efficient, zhao2015integrating, salehi2015word, vylomova2015take}, Paragraph vectors, or doc2vec, were proposed in 2014 by Quoc Le and Tomas Mikolov \cite{le2014distributed} as a simple extension to word2vec to extend the learning of embedding from words to word sequences. Doc2vec\footnote{The term doc2vec was popularized by Gensim, a widely-used implementation of paragraph vectors: https://radimrehurek.com/gensim/} is agnostic to the granularity of the word sequence, and it can equally be a word n-gram, sentence, paragraph, or document. One of the benefits of using dense and low-dimensional vectors is computational. The main benefit of this dense representation is generalization power. If we believe some features may provide similar clues, it is worthwhile to give a representation that can capture these similarities. We used text embedding to learn company names semantics and to measure similarity.

    \section{Method}\label{method}
        The method described below deals with CEM by creating a generic open solution that matches company names. Unlike many other existing solutions that utilize expensive commercial datasets, the presented method's novelty uses job-ads data, and it requires no company's information for matching besides its name. 
        
        Our method consists of two main parts: first, the creation of pairs, in which a pair consists of synonyms for the same company based on an analysis of large-scale job ads data (see Section~\ref{method-hiring company names fingerprinting}); second, the algorithm CompanyName2Vec for constructing an embedding model that utilizes these pairs of synonyms. The embedding model takes as input a company name and returns as output meaningful vector representation, where two company names that represent the same company will be relatively close vectors in terms of the Cosine distance (see Section~\ref{method-company name embedding}).

        \subsection{Hiring Company Names Fingerprinting} \label{method-hiring company names fingerprinting}
        Given a job ads corpus, we first create and enrich each job ad with a unique job ad identifier (aka job ad fingerprint) using a fingerprinting technique. 
        Second, we use the job ad fingerprint to identify company name synonyms. We group the job ad data by job ad fingerprint and create a list of company names posted with the same fingerprint, or in other words, company names that were published within the same job ad. Then, we remove duplicated synonyms and create a list of pairs, where each pair includes company names that were posted within one or more job ads. 
        
        A particular case we need to handle is staffing and recruitment agencies, which often publish their customers' job ads anonymously and use the staffing or recruitment agency name instead. 
        Ideally, each pair includes company names that refer to the same company. Therefore, to reduce such noise, we filter out staffing and recruitment agencies' names using naive text matching. Our analysis found that much of the noise can be reduced by filtering company names which include the following strings: ``staff,'' ``recruit,'' ``jobs,'' or ``unknown.''
        
        Given a dataset of job ads, our method's first step is to create a unique identifier (aka fingerprint) for each job ad. 
        This component calculates a job ad fingerprint based on the job description. We used a local approach to calculate fingerprints \cite{schleimer2003winnowing}. First we calculated a Winnowing list of fingerprints. We used the following parameters: $kgram\_len=4$, $window\_len=5$, $base=10$, and $modulo=1000$.
        Then, we utilized the MD5 hashing algorithm \cite{rivest1992md5} to convert the list of fingerprints into a single short 128-bit job identifier to improve computation and debugging efficiency (see Table~\ref{tab:fingerprints}).
            
        \subsection{Company Name Embedding} \label{method-company name embedding}
        The purpose of collecting company synonyms from job ads is to create a dataset for the method's embedding model. An outcome of it is the creation of a large database of company synonyms. We utilize the fingerprints to group company names with the same job ad's fingerprint to construct training and testing sets. After grouping the company names by fingerprint, it transforms the group of names into a set of pairs $(i_{name},j_{name})$ where $i_{name}$'s and $j_{name}$'s Job Ads have at least one Job Ad's fingerprint in common.

        It is important to note that a company may have several fingerprints, at least as many as the number of open positions it tries to fill.
        
        \begin{table*}[ht]
            \centering
            \caption{An example to four fingerprint values and the lists of hiring company names posted with the same job ad's fingerprint.}
            \begin{tabular}{l|p{8cm}}
                \textbf{Fingerprint} & \textbf{Company Names}\\
                \hline\hline
                9abc97a34a 7c5630b0673083fb9c61c1 & 99 Cents Only Stores; 99 Cent Only; 99 Cents Only; 99 Cents Only Stores LLC
                A.O. Smith; AO Smith; AO Smith Corporation; A. O. Smith; A. O. Smith Corporation; A. O. smith; Smith (A.O.) Corporation.
                \\
                \hline
                c9245756a0edca343c96a1a3b8762fc0 & ADP; ADP Automatic Data Processing; ADP Technology Services, Inc.; ADP.com; Automatic Data Processing, Inc.; ADP, Inc. \\
                \hline
                251dc200b096aaa160b744b7b907b9cc & BD; Becton Dickinson; BD (Becton, Dickinson and Company); Becton Dickinson \& Company; Becton Dickinson and Company; Becton, Dickinson \& Co.; Becton, Dickinson \& Co. (BD); Becton, Dickinson \& Company; Becton, Dickinson and Company; Becton, Dickinson; Beckon Dickinson; Beckton Dickinson. \\
                \hline
                9cc691a26109a7474efbe7c0a6f8f066 & Coca-Cola; Coca-Cola Bottling; Reyes Coca Cola Bottling; Reyes Coca-Cola Bottling; Reyes Coca-Cola Bottling Group; Coca-Cola Bottling Co. Consolidated; Coca-Cola Bottling Co. Consolidated (CCBCC); Coca-Cola Bottling Company; Coca-Cola Bottling Company Consolidated; Coke consolidated.
             
            \end{tabular}
            
            \label{tab:fingerprints}
        \end{table*}

        In addition, we removed duplicated pairs from this extensive list of names pairs, which created when two company names have more than just a single job ad fingerprint in common.
        
        In the algorithm's second step, we utilize an embedding algorithm (see Section~\ref{related work - embedding}) to capture the semantics of company names by placing semantically similar company names close together in the embedding space. We named this embedding solution CompanyName2Vec.  
        Since company names usually use minimal text, a few words, or a few dozens of characters on average, we first split the company name into characters. We used this representation for a company name instead of a word granularity level.
        Then used n-gram tokenizer, where $n$ between 1 and 3, and embedded its result using a hash function. Next, we build a sequence embedding sub-model based on a long short-term memory (LSTM) encoder \cite{greff2016lstm} with Rectified Linear Unit (ReLU) activation function \cite{agarap2018deep}. To increase the context available to the algorithm, we used a bidirectional LSTM (Bi-LSTM), a sequence processing model that consists of two LSTMs: one takes the input in a forward direction, and the other in a backward direction. We set character embedding dimensions to 400 and sequence encoder dimensions to 400 as well.
        
        We split the company synonyms into two mutually exclusive sets, train and test sets, with a ratio of 9:1, where the train set consists of about 90\% of the synonyms, and the test set 10\%. To avoid biasing of the experimental study results (see Section \ref{experimental-study-results}), we filtered out about 500 names from both the train and the test sets names, which also exist in the dataset we used for measuring the method performance - fortune 1,000 companies dataset (see Section \ref{Experimental Study - dataset}).
        
        Lastly, we calculated the embedding representation for all company names and synonyms, so it will be possible to calculate using distance metrics (see Section~\ref{related work - Distance Metrics}) the most similar names for each company name.

    \section{Experimental Study} \label{Experimental Study}
        
        There are tens of millions of open job ads in the United States published in many job boards and job search engines, like Indeed, ZipRecruiter, Glassdoor, Bing Jobs, and Google Jobs. In this study, we used a job-ads corpus from one of the biggest job boards in the United States, with more than ten million job ads. Although the dataset was acquired from a single source, there was considerable inconsistency in some jobs' properties, such as city names. For example, Los Angeles in California, USA, can be found in many variations, mainly due to letter capitalization, punctuation, and abbreviation, such as ``Los-Angeles, CA''; ``la, ca''; ``los angeles california,'' and more. Company names are also being regulated and need to comply with the state naming rules in the states where they will be doing business. Otherwise, a state might not accept the documents filed to form or qualify the company. Naming conventions include suffixes like ``LLC,'' ``Inc.,'' ``Ltd.,'' etc.

        \subsection{Dataset} \label{Experimental Study - dataset}
        According to the U.S. Bureau of Labor Statistics (BLS), there are several million job openings in the United States, which increased to a new level of almost ten million job openings after the COVID-19 pandemic hit the markets recently in 2020 \cite{bls2021trends}. However, looking at job boards and job search engines, like Indeed,\footnote{www.indeed.com} ZipRecruiter,\footnote{www.ziprecruiter.com} Glassdoor,\footnote{www.glassdoor.com} Bing Jobs,\footnote{www.bing.com} and Google Jobs,\footnote{www.google.com} show there are tens of millions of open job ads in the United States published, an order of magnitude larger than what the BLS states. We collected from one of the largest job boards in the United States a large job ads dataset that contains active job ads that existed during June 2020. This dataset includes about ten million job ads. We captured for each job ad the following information: Title, Description, Hiring Company Name, and Location (Country, State, City, and Zip Code).
        
        To evaluate the performance of the described solution, we manually created a ground-truth dataset based on the Fortune 1000 companies in the United States, where each company has a canonical company name and a list of synonyms. The mean number of synonyms for a canonical company name is 3.8, with a standard deviation of 2.1. This list consists of the most prominent American companies ranked by revenues, compiled and published by Fortune magazine \cite{furtunemagazine}. It only includes companies that are incorporated or authorized to do business in the United States and for which revenues are publicly available regardless of whether they are public companies listed on a stock market.

        \subsection{Evaluation Process and Performance Metrics} \label{Evaluation Process and Performance Metrics}
        The performance of the recommendations generated by our method was measured by the average of $Success@k$ (Success at rank k) rate, which stands for the mean probability that a relevant company name occurs within the top k of the ranking \cite{garcin2014offline}.  
        Similar to how the relevance of the search engines is measured, in most cases, people are only interested in the first page of the results and do not bother to move on to subsequent pages \cite{brin1998anatomy}. In extreme cases, only the top result matters. For example, only a single value can be used when performing a join operation between two tables in a database.
        
        We compared the solution's performance with edit-distance matching, fuzzy distance matching, and random matching, common best algorithms, as described in Section~\ref{related work}. 
        We treated the ground-truth dataset (see Section~\ref{Experimental Study - dataset}) as a mapped list of synonyms to tags, where a tag is the most common company name (canonical name). 
        
        We used the ground-truth dataset with the created model described in Section~\ref{method} and generated a vector representation for both canonical names and synonyms. For each synonym's vector, we calculated its cosine distance with all the canonical name's vectors. As a result, we generated a $N\times M$ matrix $R$, where $N$ is the total number of canonical names in the ground-truth dataset, and $M$ is the size of the complete set of synonyms. R$_{ij}$ represents the cosine distance between the vector representation of a canonical name $i$ and the vector representation of a synonym $j$.
        For each given synonym, we sorted the canonical names based on their cosine distance $R_{ij}$. Lastly, the closest $k$ canonical names were presented as the recommended canonical names for each given synonym. We used the $k$ closest vectors provided in the matrix $R$, and calculated the metric of $Success@k$, for $k = 1, 2, 3$. We defined success as equal to 1 in case the canonical company name returned in the closest $k$ company names, or to be equal to 0 otherwise. Namely, we define $Success@k$ as follows:

        \begin{equation*}
        \begin{split}
        Success@k(synonym) := &|\{Tagged \ Canonical \ Name\} \cap  \\
        & \{k \ Closest \ Canonical \ Names \}|    
        \end{split}
        \end{equation*}
        
        Where tagged canonical company name is mapped to the company synonym in the ground-truth dataset. 
        
        \begin{equation*}
        Avg\-Success@k := \frac {\sum_{s \in Synonyms} success@k(s)} {|Synonyms|}
        \end{equation*}

    \section{Results} \label{experimental-study-results}
        To analyze the performance of the implemented method (see Section~\ref{method}), we used the job ads dataset (see Section~\ref{Experimental Study - dataset}) to train the embedding CompanyName2Vec model (see Section~\ref{method-company name embedding}), and evaluated the method's performance using the ground-truth dataset (see Section~\ref{Experimental Study - dataset}). Since the ground-truth dataset is biased to company names which oprate in the United States, we first checked the geographical distribution of job corpus dataset. We found that the majority of the job ads, about $88.5\%$, are posted within the United States. 
        
\begin{figure}[ht]
            \centering
            \includegraphics[width=0.7\textwidth]{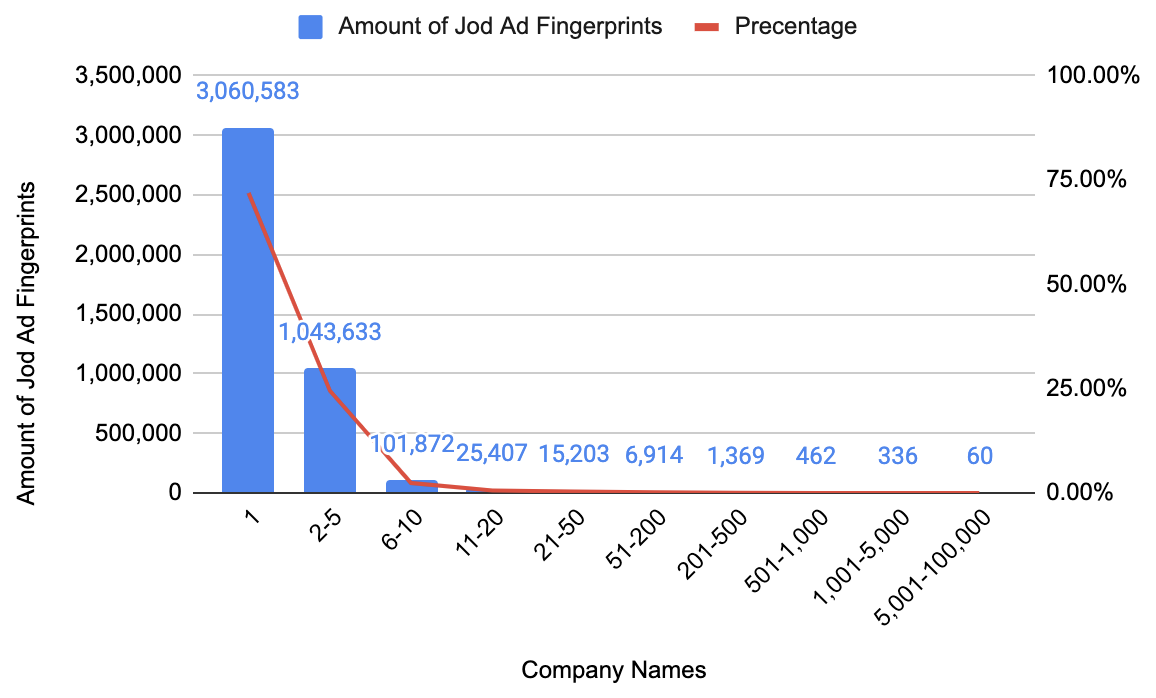}
            \caption{Fingerprints Distribution by Company Names.}
            \label{fig:fingerprints-distribution-by-company-names-before-filtering}
        \end{figure}

        The job ads fingerprinting result (see Section~\ref{method-hiring company names fingerprinting}) shows that there are 4.25 million of unique job ad fingerprints out of 10 million job ads. From the distribution of job ads fingerprints (see Figure~\ref{fig:fingerprints-distribution-by-company-names-before-filtering}), only 1.2 million of the unique job ad fingerprints, which were posted with at least two different company names, could be used for training the embedding CompanyName2Vec model. We found that applying the agencies staffing companies filters (see Section~\ref{method-hiring company names fingerprinting}) removed 9.5\% of the job ads, 8.25\% of the fingerprints, and just 2.5\% of the company names.
        
\begin{figure*}
            \centering
            \includegraphics[width=1\textwidth]{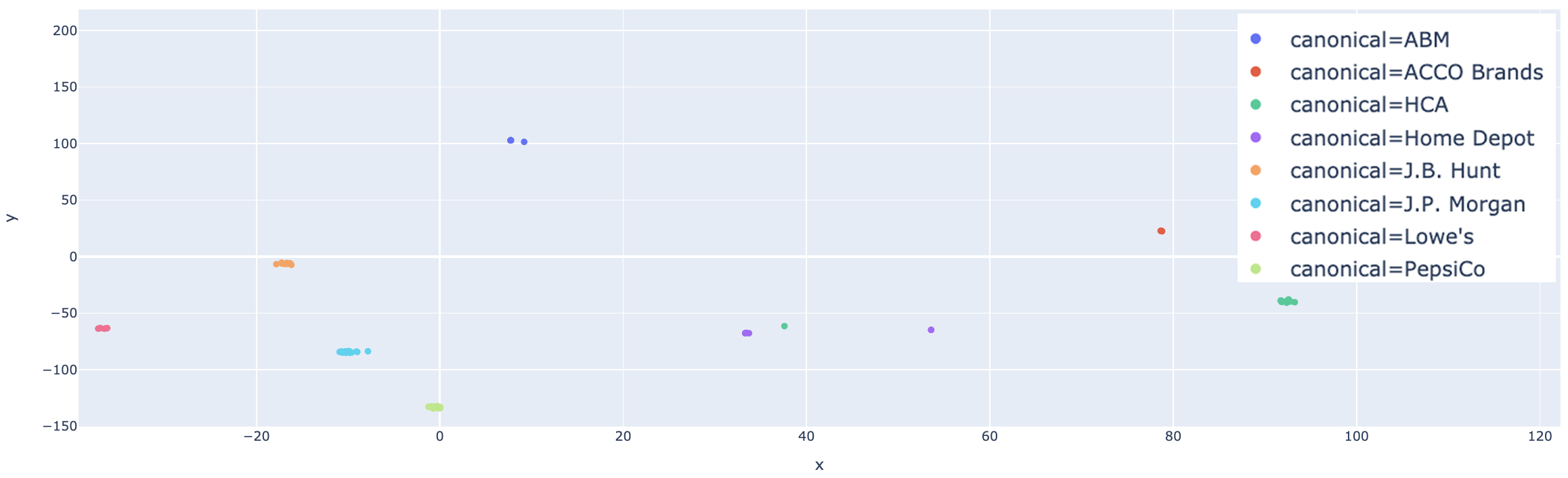}
            \caption{Synonym distribution of the following companies as were calculated by the CompanyName2Vec algorithm and reduced into two dimensions with t-SNE: ABM Industries, ACCO Brands, HCA Healthcare, Home Depot, J.B. Hunt Transport Services, Inc., JPMorgan Chase \& Co., Lowe's Inc., and PepsiCo.}
            \label{fig:synonyms distribution}
        \end{figure*}

        We generated the embedding CompanyName2Vec model based on the job ads dataset (see Section~\ref{Experimental Study - dataset}). We used the generated model for resolving CEM and matched the Fortune 1000 canonical company names with their synonyms. Figure~\ref{fig:synonyms distribution} presents the distribution of the associated names of the following companies based on the CompanyName2Vec vector representation after being reduced into two dimensions using t-SNE \cite{van2008visualizing}: ABM Industries, ACCO Brands, HCA Healthcare, Home Depot, J.B. Hunt Transport Services, Inc., JPMorgan Chase \& Co., Lowe's Inc., and PepsiCo.
        As can be seen, there are eight clusters, one per every company includes the company synonyms grouped together, which indicates that CompanyName2Vec successfully suggested relevant synonyms for those companies.

        Lastly, we calculated $AvgSuccess@k$ (see Figure~\ref{Evaluation Process and Performance Metrics}) for the CompanyName2Vec solution and compared it with the following several solutions: random matching, edit-distance matching \cite{navarro2001guided}, and fuzzy matching \cite{chaudhuri2005robust}. The CompanyName2Vec performs better than all the above solutions, with any given tested $k$ (see Table~\ref{tab:Solutions' Performance Comparison}). We evaluated this list of algorithms using a 10-fold cross-validation approach. The difference found statistically significant using
        t-tests with p-value less than 0.001.

        \begin{table}[ht]
            \centering
            \caption{Algorithms' Performance Comparison}
            \includegraphics[scale=0.48]{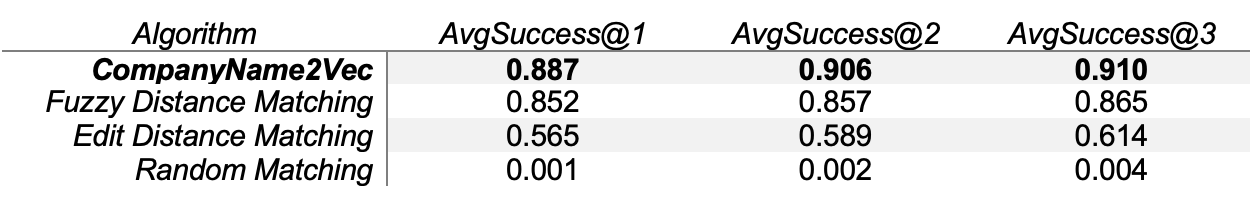}
            \label{tab:Solutions' Performance Comparison}
        \end{table}

        \begin{figure}
            \centering
            \begin{subfigure}{1\textwidth}
                \centering
                \includegraphics[width=1\textwidth]{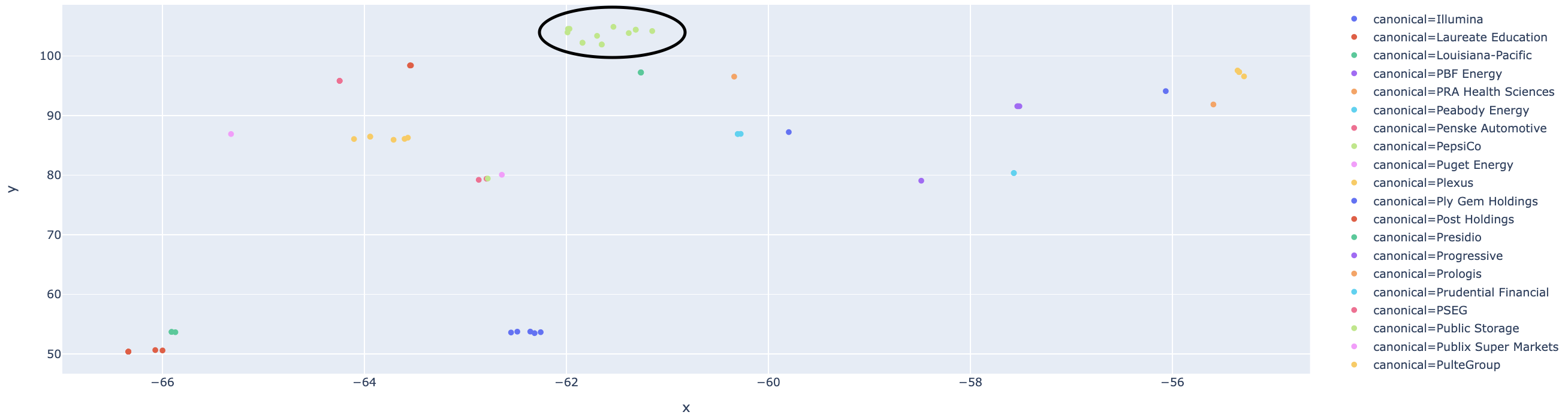}
                \caption{Zoom out view on PepsiCo synonyms and other company synonyms near by.}
                \label{fig:pepsico plot - PepsiCo comparison}
            \end{subfigure}
            \hfill
                \begin{subfigure}{1\textwidth}
                \centering
                \includegraphics[width=1\textwidth]{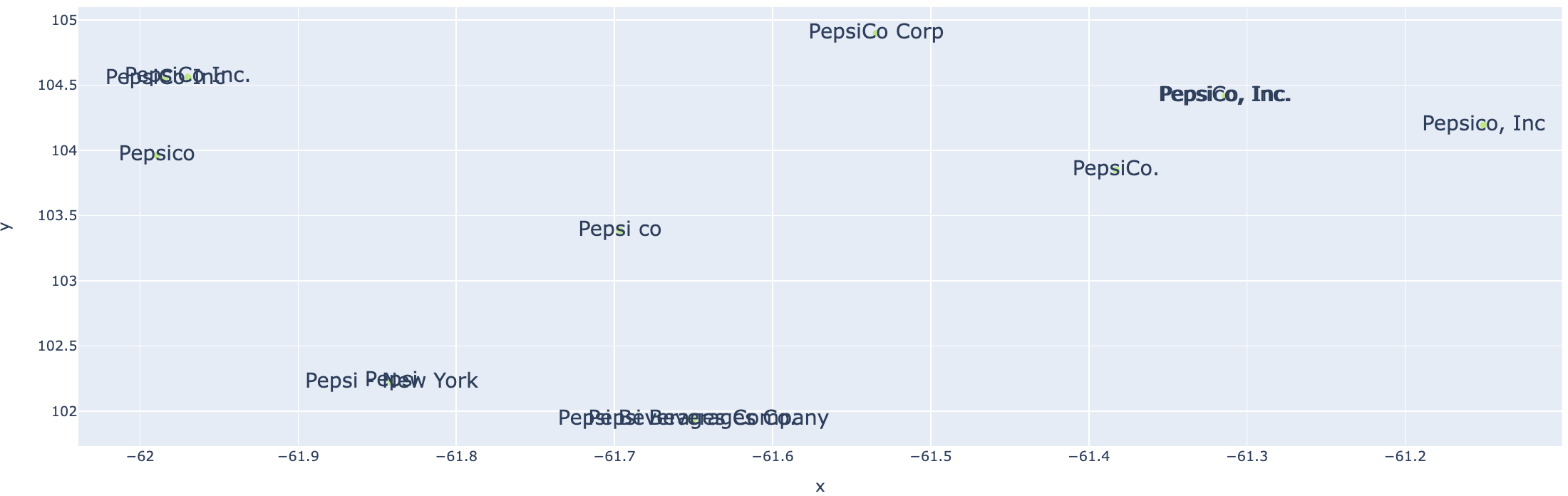}
                \caption{Examples for PepsiCo synonyms}
                \label{fig:pepsico plot - PepsiCo synonyms}
            \end{subfigure}
            \caption{A plot of PepsiCo synonyms' vectors after were reduced into two dimensions.}
            \label{fig:pepsico plot}
        \end{figure}

        \begin{figure}
            \centering
            \begin{subfigure}{1\textwidth}
                \centering
                \includegraphics[width=1\textwidth]{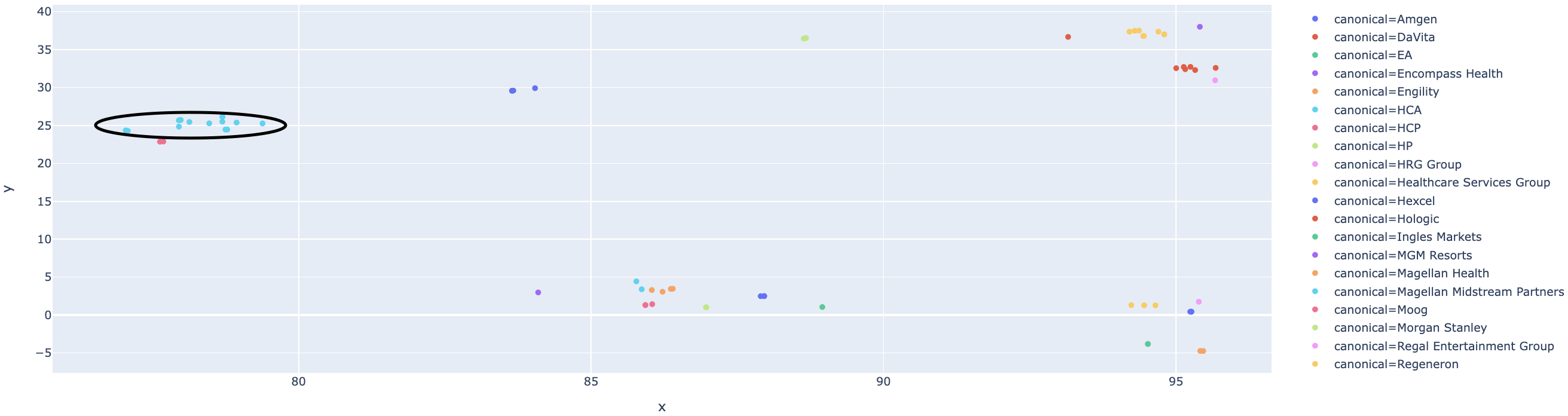}
                \caption{Zoom out view on HCA Healthcare synonyms and other company synonyms near by.}
                \label{fig:hca plot - hca comparison}
            \end{subfigure}
            \hfill
                \begin{subfigure}{1\textwidth}
                \centering
                \includegraphics[width=1\textwidth]{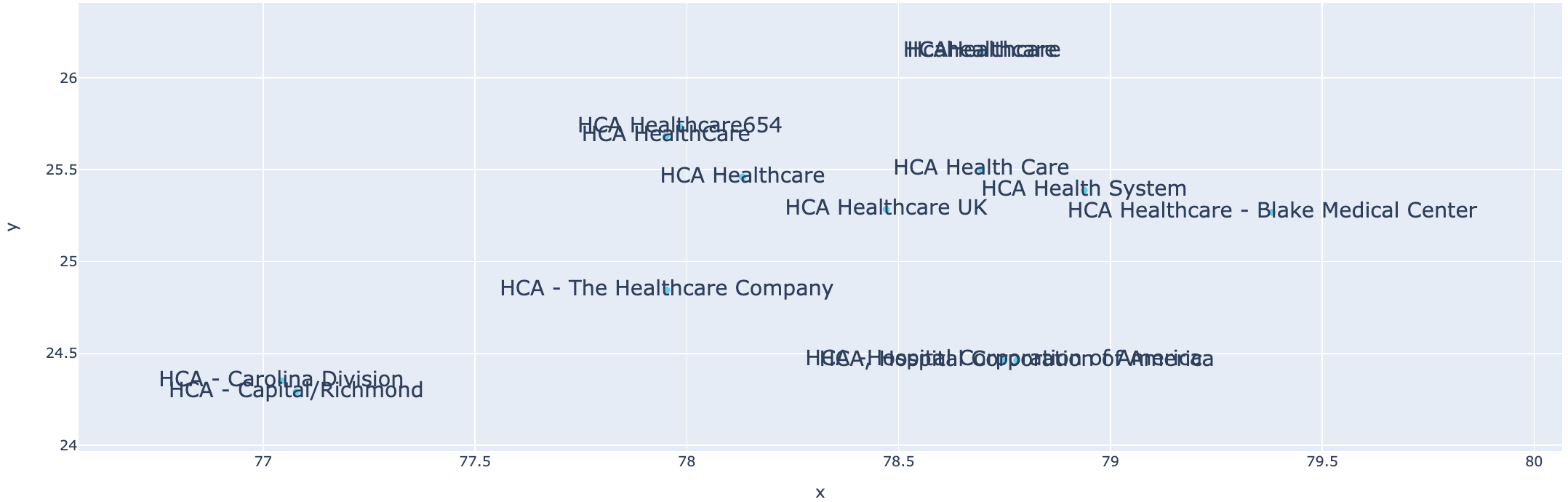}
                \caption{Examples for HCA Healthcare synonyms}
                \label{fig:hca plot - hca synonyms}
            \end{subfigure}
            \caption{A plot of HCA Healthcare synonyms' vectors after were reduced into two dimensions.}
            \label{fig:hca plot}
        \end{figure}

        \begin{figure}
            \captionsetup{justification=centering}
            \centering
            \begin{subfigure}{0.48\textwidth}
                \centering
                \includegraphics[width=\textwidth]{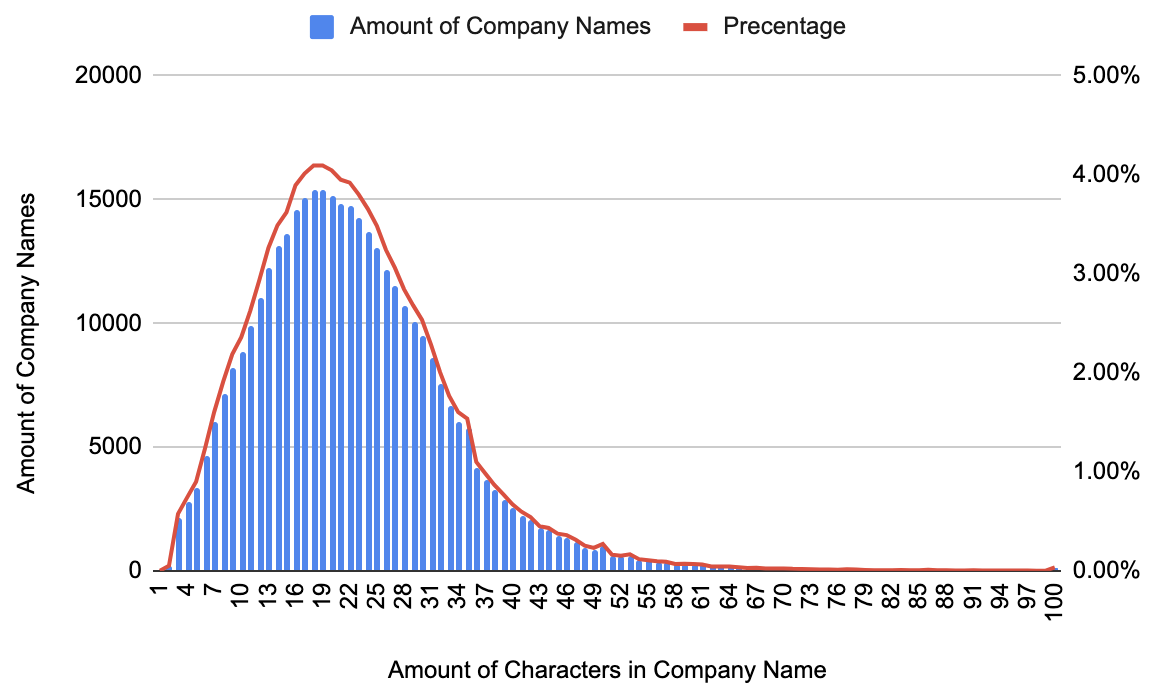} 
                \caption{Distribution of Company Names by Name's Length}
                \label{fig:Amount of Company Names vs. Company Name Length}
            \end{subfigure}
            \hfill
            \begin{subfigure}{0.48\textwidth}
                \centering
                \includegraphics[width=\textwidth]{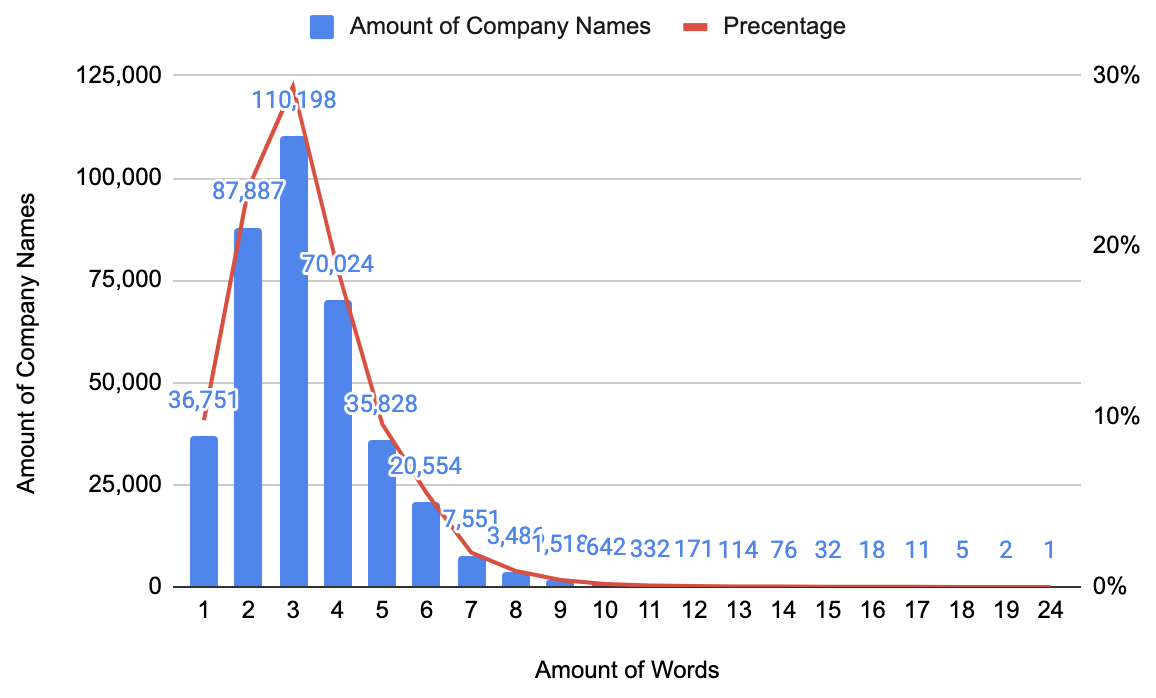}
                \caption{Distribution of company names by amount of words}
                \label{fig:Amount of Company Names vs. Amount of Words}
            \end{subfigure}
            \hfill
                \caption{Distribution of company names exist in job ad corpus}
                \label{fig:Company Names Length}
        \end{figure}

    \section{Discussion}\label{sec:discuss}
        Upon analyzing the results presented in the previous section,  we can conclude the following:
        First, the novel approach we call CompanyName2Vec, which defines a vector representation of company names based on job ads, had promising results compared to other algorithms (see Table~\ref{tab:Solutions' Performance Comparison}) and is helpful for the task of CEM (see Section~\ref{Evaluation Process and Performance Metrics}). CompanyName2Vec algorithm captures the company names semantics and manages common prefixes, suffixes, and punctuation for company synonyms, like PepsiCo's suffixes - Inc, Corp., Co., Company, Etc. (see Figure~\ref{fig:pepsico plot}). CompanyName2Vec manages less popular suffixes as well, which may include a location or a point of interest. For example, HCA Capital Richmond is a synonym of HCA Healthcare. HCA Capital Richmond is a synonym due to the presence of HCA capital offices, which are located in Richmond, VA (see Figure~\ref{fig:hca plot}). Another example is \textit{Pepsi - New York}, which is a synonym of PepsiCo's office in New York City (see Figure~\ref{fig:pepsico plot}).
        A company's business line can be attached to a company name as a synonym, such as Home Depot Tools Rental (see Figure~\ref{fig:home depot plot}), or Lowe's Home Improvement (see Figure~\ref{fig:lowes plot}). Home Depot Tools Rental is a synonym for Home Depot, and Lowe's Home Improvement is the company slogan.
        
        Second, we observed that ``best partial'' heuristic (see Section~\ref{Character-Based Similarity Metrics}) works well in the case of matching company names because of its short length. The median length of a company name is 21 characters (see Figure~\ref{fig:Amount of Company Names vs. Amount of Words}), and it consists in the median case of just three words (see Figure~\ref{fig:Amount of Company Names vs. Amount of Words}). 
        
        Third, based on the method's performance comparison (see Table~\ref{tab:Solutions' Performance Comparison}), we found the CompanyName2Vec outperforms the other algorithms tested, while the second-best algorithm measured is the Fuzzy distance matching. 
        
        Fourth, as for company abbreviations, due to lack of semantics in company name abbreviations, the CompanyName2Vec solution performs just as well as Fuzzy distance matching.
        
        Lastly, we found the CompanyName2Vec method works better than the Fuzzy method mainly because the CompanyName2Vec uses the language semantic. For example, trying to match ``Pepsi - New York" with ``PepsiCo Inc.'' and ``New York Life'' using Fuzzy distance matching returns ``New York Life'' because New York is a significant part of both names. The CompanyName2Vec method returns ``PepsiCo Inc." based on an LSTM model, which extracts the company naming semantics from the training set.

    \section{Conclusions and Future Work} \label{sec:conclusions}
        In this study, we introduced CompanyName2Vec, a novel, generic open-source algorithm which uses job ads and deep learning to address some of the challenges associated with company name synonyms. 
        Another contribution of this study is an end-to-end, highly scalable, enterprise-grade system that uses the CompanyName2Vec algorithm to solve the CEM problem using job-ads data and requires no company's information for matching besides its name.
        We provided a comprehensive description of our algorithm's steps, starting with collecting job ads and finding company name synonyms using the job ads fingerprinting technique and generating the company name synonyms dataset. We used text matching heuristics to reduce wrong synonyms and used the dataset to generate a Bi-LSTM model. This model generates a vector representation given any company name. We collected a set of synonyms for the Fortune 1000 companies in the United States and used them to test and compare the CompanyName2Vec algorithm with several matching algorithms. We matched all the Fortune 1000 companies' synonyms with their canonical names and analyzed the results.  
        The evaluation demonstrated that CompanyName2Vec was beneficial for confronting the problem of recommending synonyms for a given company name. CompanyName2Vec performed best out of the other evaluated methods with the higher $Success@k$ for any tested $k$ value. 
        
        Our research uses the job ads' description and the Winnowing algorithm for job ad fingerprinting. A possible future research direction is using a more comprehensive job ads' fingerprinting technique using additional information than just the job description, such as work location, contact information, company logo, Etc.
        Another possible research direction can be exploring other models than Bi-LSTM for name embedding, like bag-of-words, convolutional neural network (CNN), and bidirectional encoder representations from transformer (BERT). 
        Another avenue to pursue is learning company relations from the job ads corpus, for example, a parent company, a subsidiary, a holding company, etc.

    \bibliography{main}
    
	\appendix
	\setcounter{section}{0}
	\setcounter{figure}{0}
\renewcommand{\thesection}{\Alph{section}}
\renewcommand\thefigure{S\arabic{figure}}
\section{Appendix}
	
	    \begin{figure}
            \centering
            \begin{subfigure}{1\textwidth}
                \centering
                \includegraphics[width=1\textwidth]{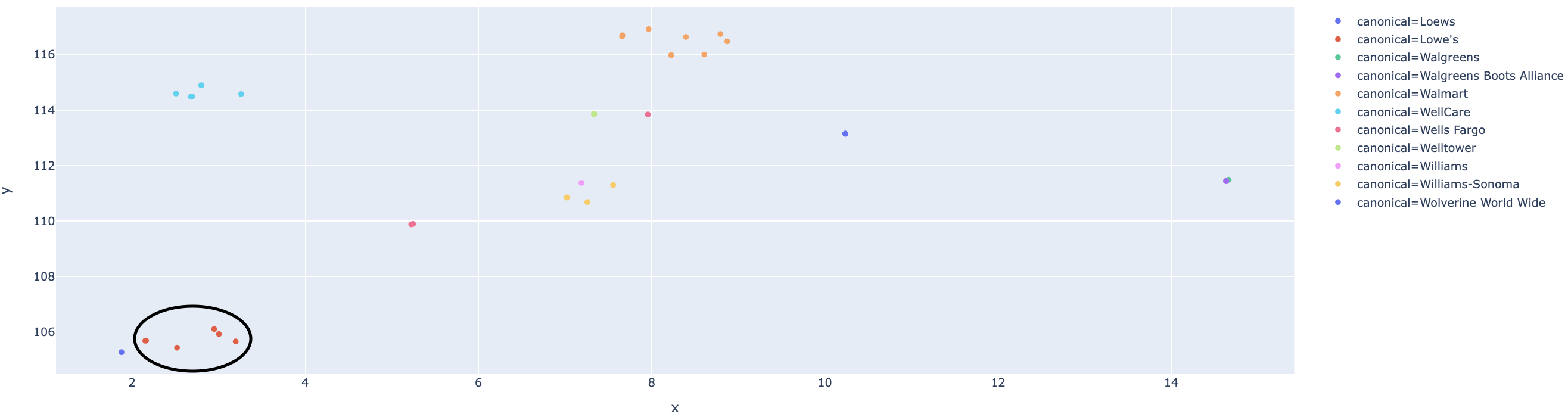}
                \caption{Zoom out view on Lowe's Inc. synonyms and other company synonyms near by}
                \label{fig:lowes plot - lowes comparison}
            \end{subfigure}
            \hfill
            \begin{subfigure}{1\textwidth}
                \centering
                \includegraphics[width=1\textwidth]{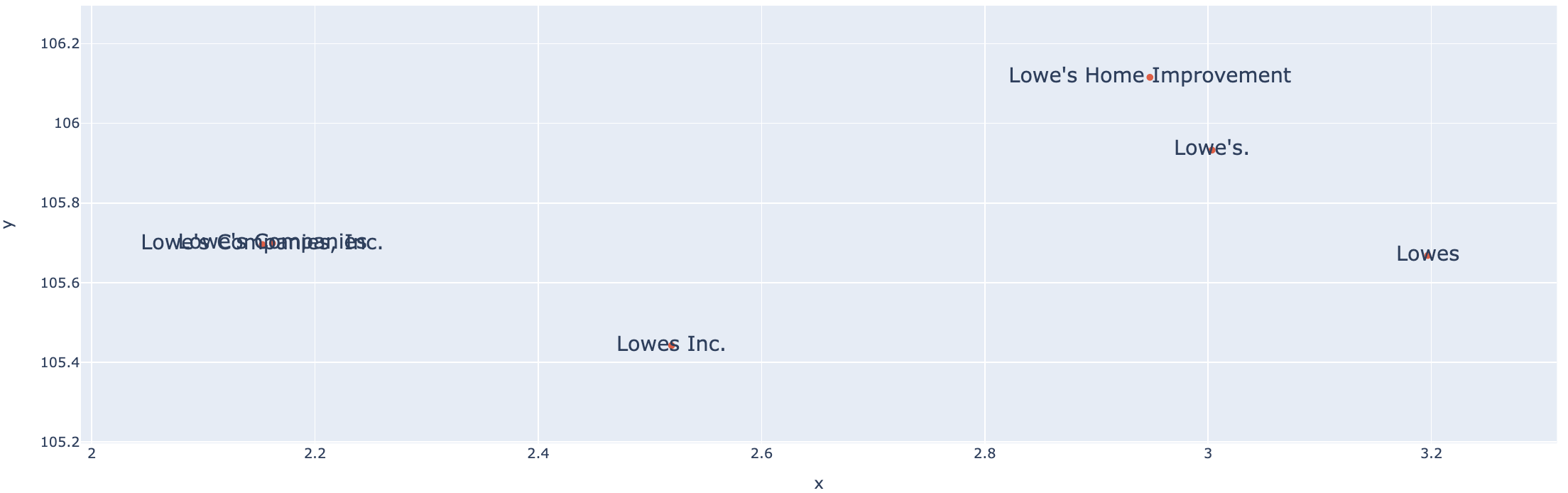}
                \caption{Examples for Lowe's Inc. synonyms}
                \label{fig:lowes plot - lowes synonyms}
            \end{subfigure}
            \caption{A plot of Lowe's Inc. synonyms' vectors after were reduced into two dimensions}
            \label{fig:lowes plot}
        \end{figure}

        \begin{figure}
            \centering
            \begin{subfigure}{1\textwidth}
                \centering
                \includegraphics[width=1\textwidth]{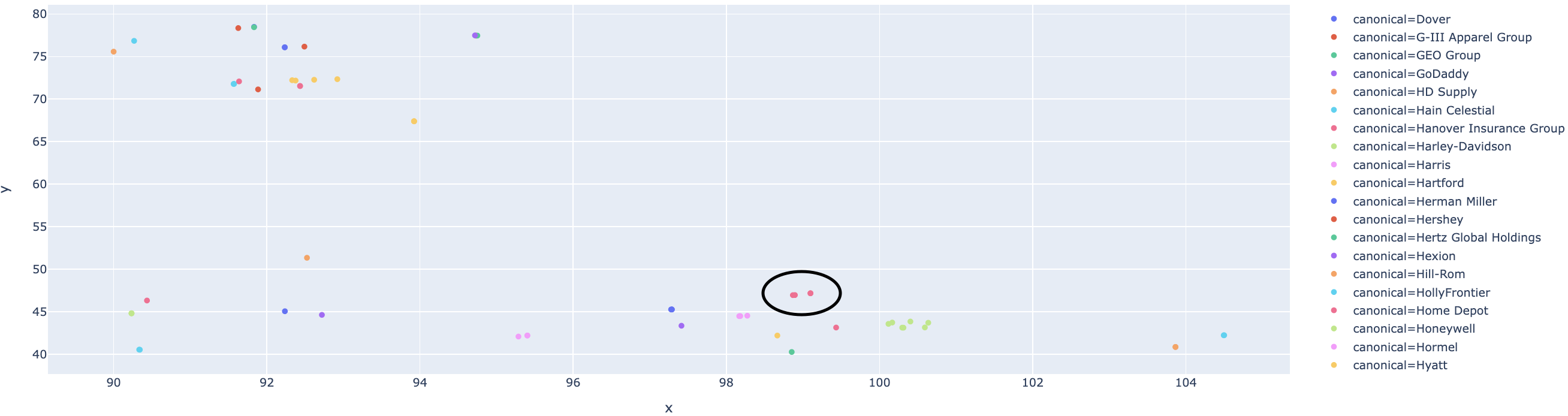}
                \caption{Zoom out view on Home Depot synonyms and other company synonyms near by}
                \label{fig:home depot plot - home depot comparison}
            \end{subfigure}
            \hfill
                \begin{subfigure}{1\textwidth}
                \centering
                \includegraphics[width=1\textwidth]{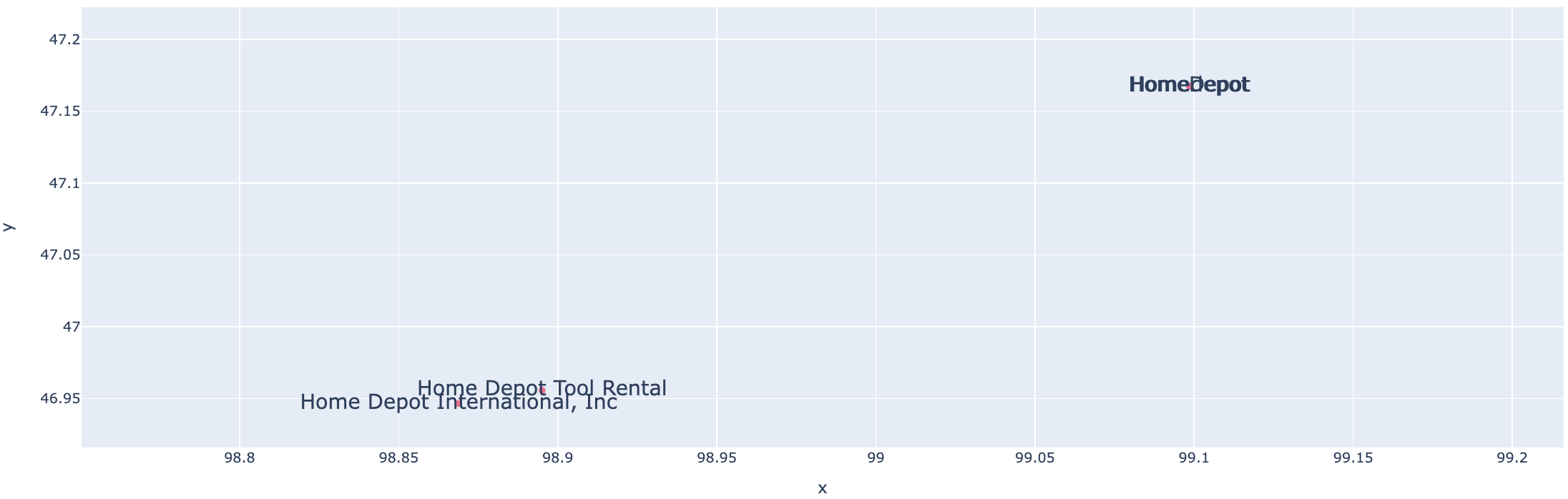}
                \caption{Examples for Home Depot synonyms}
                \label{fig:home depot plot - home depot synonyms}
            \end{subfigure}
            \caption{A plot of Home Depot synonyms' vectors after were reduced into two dimensions}
            \label{fig:home depot plot}
        \end{figure}

\end{document}